\pdfoutput=1
\documentclass[aps,prb,twocolumn,superscriptaddress,
nopreprintnumbers,amsmath,amssymb]{revtex4}
\usepackage{graphicx,subfigure}
\usepackage{dcolumn}
\usepackage{bm}
\usepackage{ulem}
\usepackage{color}
\definecolor{orange}{rgb}{1,0.5,0}

\begin{document}

\title{Kitaev-Heisenberg-$J_2$-$J_3$ model for the iridates $\text{A}_2\text{Ir}\text{O}_3$}
\author{Itamar Kimchi}
\affiliation{Department of Physics, University of California, Berkeley, CA 94720, USA}
\author{Yi-Zhuang You}
\affiliation{Department of Physics, University of California, Berkeley, CA 94720, USA}
\affiliation{Institute for Advanced Study, Tsinghua University, Beijing, 100084, China}
\date{\today}
\begin{abstract}
A Kitaev-Heisenberg-$J_2$-$J_3$ model is proposed to describe the Mott-insulating layered iridates A$_2$IrO$_3$ (A=Na,Li). The model is a combination of the Kitaev honeycomb model and the Heisenberg model with all three nearest neighbor couplings $J_1$, $J_2$ and $J_3$. A rich phase diagram is obtained at the classical level, including the experimentally suggested zigzag ordered phase; as well as the stripy phase, which extends from the Kitaev-Heisenberg limit to the $J_1$-$J_2$-$J_3$ one. Combining the experimentally observed spin order with the optimal fitting to the uniform magnetic susceptibility data gives an estimate of possible parameter values, which 
in turn reaffirms the necessity of including both the Kitaev and farther neighbor couplings.
\end{abstract}
\maketitle
\begin{figure}[tbhp]
\begin{center}
\includegraphics[height=0.11\textheight]{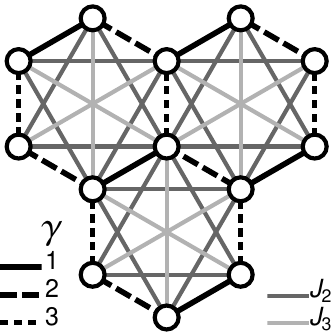}
\caption{The honeycomb lattice, with Kitaev label $\gamma$ for nearest neighbor bonds, and including second and third neighbor bonds with Heisenberg couplings $J_2$ and $J_3$.}
\label{fig:lattice}
\end{center}
\end{figure}

Frustrated spin systems have long served as a relatively simple yet rich source of exotic phenomena such as spin liquids and unconventional order.
 The frustration may arise either geometrically on a lattice incompatible with the spin ordering, or dynamically from non-commuting competing terms in the Hamiltonian. The nearest neighbor $S=1/2$ Heisenberg model on the kagome lattice is an instance of geometrical frustration that may even host a quantum spin liquid ground state\cite{kagome}. 
Bipartite lattices such as the honeycomb can still be geometrically frustrated by including farther than nearest neighbor antiferromagnetic Heisenberg exchange, giving so called $J_1$-$J_2$-$J_3$ models. Such models on the honeycomb in particular have seen a recent surge of work\cite{Albuqu, Fouet, PFRGj123,   FarnellPVBC,  OitmaaSingh}, though a quantum spin liquid phase may require charge as well as spin fluctuations\cite{MengNature, AbaninSondhi, YangSchmidt}. Breaking spin rotational symmetry provides avenues for dynamical frustration, as in the Kitaev honeycomb model\cite{Kitaev}, a nearest neighbor Ising coupling of spin component set by a bond label $\gamma$ as in Fig.\,\ref{fig:lattice}.  This seemingly artificial model is exactly solvable with a spin liquid ground state exhibiting an emergent Majorana fermion with a $Z_2$ gauge background. 

\begin{figure}[tbhp]
\begin{center}
\subfigure[]{\includegraphics[height=0.155\textheight]{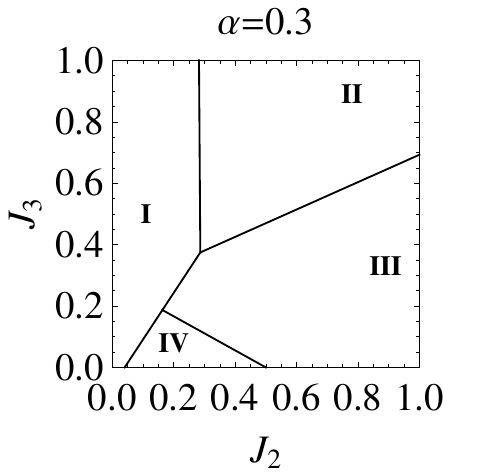}\label{fig:alpha3phases}}
\\[-8 pt]
\subfigure[]{\includegraphics[height=0.09\textheight]{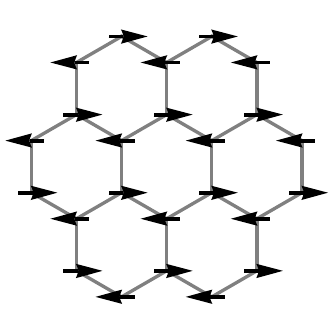}\label{fig:afconfig}}
\subfigure[]{\includegraphics[height=0.09\textheight]{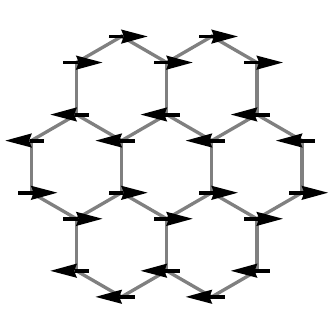}\label{fig:zzconfig}}%\qquad
\subfigure[]{\includegraphics[height=0.09\textheight]{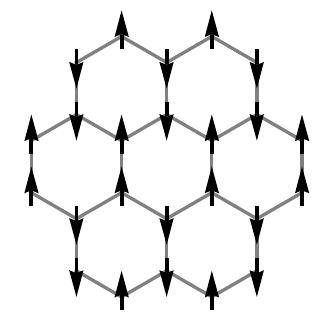}\label{fig:stconfig}}
\caption{(a) Sample ($J_2,J_3$) slice of the classical phase diagram, with phases (I), (II) and (IV) represented in (b), (c) and (d) respectively. Region (III) contains various noncollinear spiral configurations. \ (b) (I) \textit{Neel}. (c) (II) \textit{Zigzag}. (d) (IV) \textit{Stripy}.}
 \label{fig:configs}
\end{center}
\end{figure}

A recent and surprising addition to the experimentally relevant $J_1$-$J_2$-$J_3$ models of frustrated spin systems, the Kitaev coupling has been recently proposed \cite{JK,CJK} to occur in the Mott insulating\cite{Na2IrO3} iridates 
$\text{A}_2\text{Ir}\text{O}_3$
(A=Na,Li), where the iridium ions are arranged in layers of 2D honeycomb lattices.  
Uniform susceptibility and heat capacity studies on these materials\cite{Na2IrO3,Li2IrO3} found Curie-Weiss temperatures of $-125$ K for $\text{Na}_2\text{Ir}\text{O}_3$ and $-33$ K for $\text{Li}_2\text{Ir}\text{O}_3$, and a low magnetic ordering temperature of 15 K for both, suggesting strong frustration. A resonant x-ray scattering measurement\cite{zigzag} on $\text{Na}_2\text{Ir}\text{O}_3$ found the ground state has antiferromagnetic order at wavevector $M$, suggested by a first principles calculation\cite{zigzag} to be a \textit{zigzag} rather than a \textit{stripy} configuration (see Fig.\,\ref{fig:configs}). 

Strong spin-orbit coupling splits the iridium $t_{2g}$ states into a filled manifold and a half filled Kramer's doublet, an effective spin-1/2 degree of freedom which need no longer respect the rotational symmetry. Thus the $90^{\circ} $ angles of the Ir-O-Ir hopping path within the oxygen octahedra, together with d-orbital Hund's rule coupling and orbital interactions, are able to give the Kramer's doublet highly anisotropic exchanges of the Kitaev form. Higher order hopping paths, direct orbital overlaps, trigonal distortions and spin-orbit energy splittings within the iridium two electron propagator all contribute spin interactions other than the Kitaev term, primarily including antiferromagnetic Heisenberg exchange. 

Keeping only the nearest neighbor Heisenberg exchange yields the Heisenberg-Kitaev model\cite{CJK,stationQ,finiteT,HKvacancies}, Eq.\,(\ref{eq:H}) with $J_2,J_3$ set to zero, which has been previously used to describe the $\text{A}_2\text{Ir}\text{O}_3$ materials\cite{Na2IrO3,Li2IrO3,zigzag,CJK}. The phase diagram\cite{CJK,stationQ} in the parameter $0\leq \alpha \leq 1$ consists of a Neel phase for the Heisenberg model at small $0\leq \alpha<0.4$, the Kitaev spin liquid at large $0.8 < \alpha \leq 1$, and an intermediate antiferromagnetically ordered \textit{stripy} phase (see Fig.\,\ref{fig:configs}). The \textit{stripy} configuration is the exact ground state at $\alpha=0.5$, solvable by means of a periodic site dependent spin rotation\cite{CJK} which turns the Hamiltonian into a Heisenberg ferromagnet in the rotated spins.

Preserving $J_2$ and $J_3$ to produce the previously unstudied Kitaev-Heisenberg-$J_2$-$J_3$ model is important for two reasons. First, substantial $J_2$ and $J_3$ are likely to exist in the materials; density functional theory (DFT) calculations\cite{Jin} for $\text{Na}_2\text{Ir}\text{O}_3$ found $J_2 / J_1 \approx 0.5$, and a later tight binding fit of the DFT data including $J_3$ found $J_2, J_3$ to be approximately equal\cite{YBKtrig}.
Second, the experimentally suggested \textit{zigzag} ordered ground state\cite{zigzag} can not be realized in a Kitaev-Heisenberg model alone. It is found that an antiferromagnetic $J_3$ term is needed to stabilize the \textit{zigzag} order. Moderate Kitaev and $J_2$ couplings stabilize both \textit{zigzag} and \textit{stripy} orders. 
  We will also show that in order to reproduce the experimentally measured uniform susceptibility $\chi(T)$, the farther neighbor $J_2$ and $J_3$ couplings as well as the Kitaev term are likely needed.

 The Kitaev-Heisenberg-$J_2$-$J_3$ Hamiltonian is
\begin{equation}\label{eq:H}
\begin{split}
H = J&\left[(1-\alpha)\left(\sum_{\langle i j\rangle}+J_2\sum_{\langle\langle i j\rangle\rangle}+J_3\sum_{\langle\langle\langle i j\rangle\rangle\rangle}\right)\bm{\sigma }_i\cdot\bm{\sigma }_j \right. \\
&\left. -2\alpha\sum_{\langle i j\rangle } \sigma^{\gamma_{ij}}_i \sigma^{\gamma_{ij}}_j \right]
\end{split}
\end{equation}
where $\langle i j\rangle$, $\langle\langle ij\rangle\rangle$ and $\langle\langle\langle ij\rangle\rangle\rangle$ stand for the first, second and third nearest neighbor bonds, and $\gamma_{ij}$ is a nearest neighbor bond label, as illustrated in Fig.\,\ref{fig:lattice}. The model interpolates between the $J_1$-$J_2$-$J_3$ model at $\alpha=0$ and the Kitaev model at $\alpha=1$, maintaining the second and third neighbor coupling strengths $J_2$  and $J_3$ in units of the nearest neighbor Heisenberg coupling strength.

A recently proposed alternative model for Na$_2$IrO$_3$ based on \textit{ab initio} calculations\cite{Jin} takes the limit where trigonal distortion effects are stronger than spin-orbit coupling, finding a Hamiltonian with Ising anisotropy and no Kitaev term\cite{YBKtrig, pyroTrig}. 
Putting this interesting scenario aside\cite{QSHE}, we find that mild $\hat{c}\equiv (1,1,1)$ uniaxial trigonal distortion is consistent with our approach.  The effective spin-1/2 Kramer's doublet remains well separated from the filled states. 
Its modified wavefunction creates anisotropies in the magnetic field coupling ($g$-factor tensor) and combines with the non-$90^\circ$ Ir-O-Ir hopping path to perturb Eq.\,(1),
possibly enhancing both Kitaev and Heisenberg terms in addition to creating small Ising $S^{\hat{c}}S^{\hat{c}}$ and Ising-Kitaev $S^{\gamma_{ij}}S^{\hat{c}}$ terms. Both modifications are expected from the observed anisotropy  in single crystal $\text{Na}_2\text{Ir}\text{O}_3$ susceptibility\cite{Na2IrO3} and do not change our results.


Since there is ample evidence\cite{Na2IrO3,Li2IrO3,zigzag} for magnetic ordering in both $\text{Na}_2\text{Ir}\text{O}_3$ and $\text{Li}_2\text{Ir}\text{O}_3$ we will leave the calculation of the quantum phase diagram of Eq.\,(\ref{eq:H}) for future work, instead turning to the magnetically ordered phases which may be studied by a purely classical analysis.
For each point $\left( \alpha ,J_2,J_3 \right)$ in the three dimensional phase diagram we determined the magnetic ordering configurations using a quadratic (unconstrained) classical spin model\cite{Anderson}, which we diagonalized analytically in momentum space. Since $\Gamma =-\Gamma $ and $M=-M$ these two wavevectors automatically give configurations of collinear unit-length normalized spins despite the absence of the unit-length constraint in the calculation, reaffirming the validity of the classical solution at these points. Solutions at wavevector $K$ or at generic incommensurate wavevectors correspond to noncollinear spiral configurations, which we label as a single phase. 

\begin{figure*}[bthp]
\begin{center}
\subfigure[]{\includegraphics[height=125 pt]{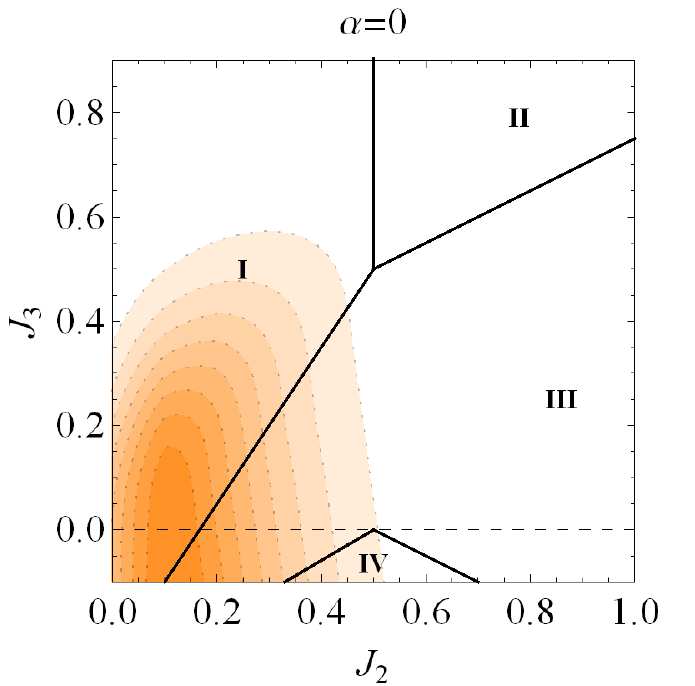}}
\subfigure[]{\includegraphics[height=125 pt]{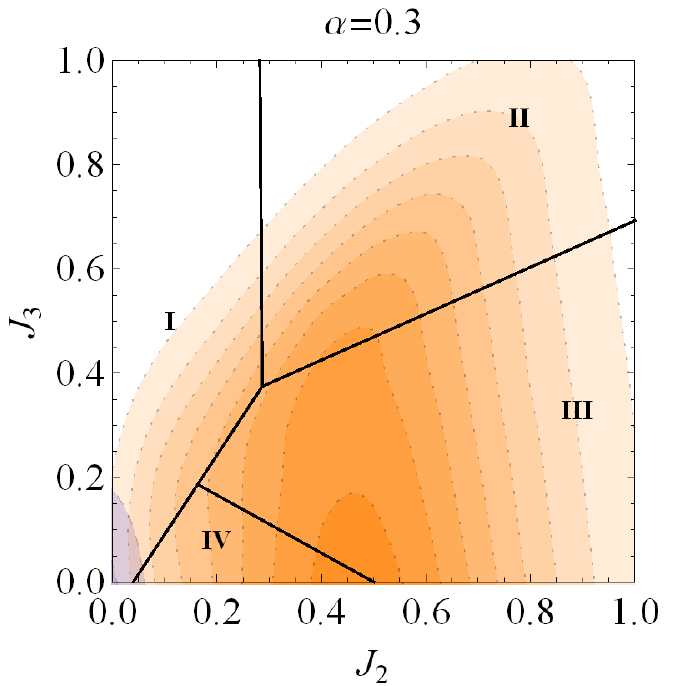}}
\subfigure[]{\includegraphics[height=125 pt]{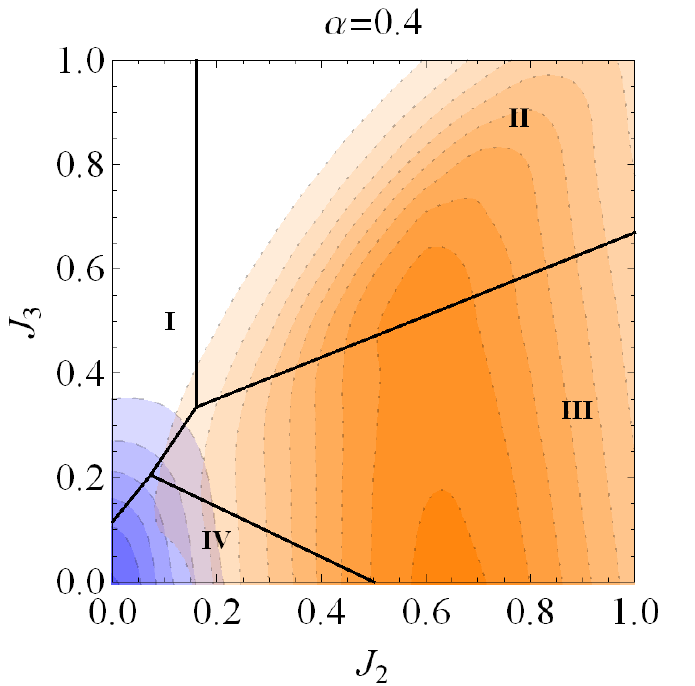}}
\\
\subfigure[]{\includegraphics[height=125 pt]{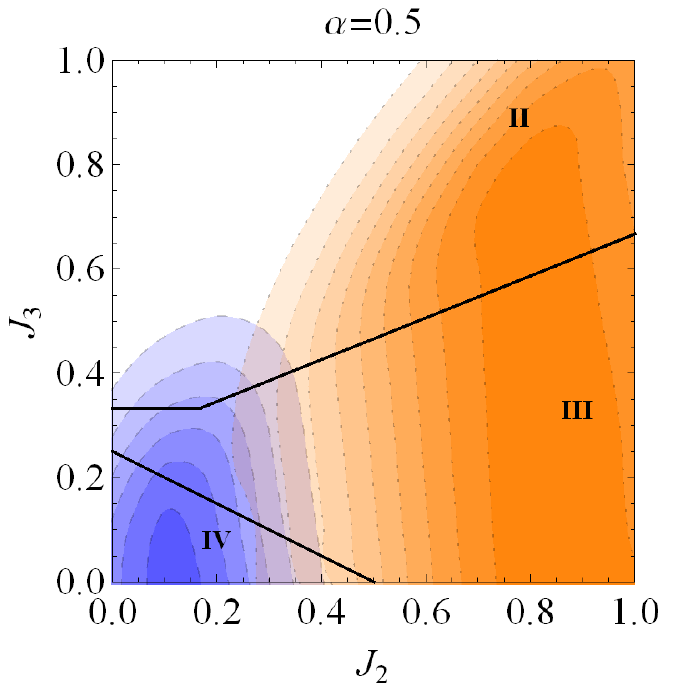}}
\subfigure[]{\includegraphics[height=125 pt]{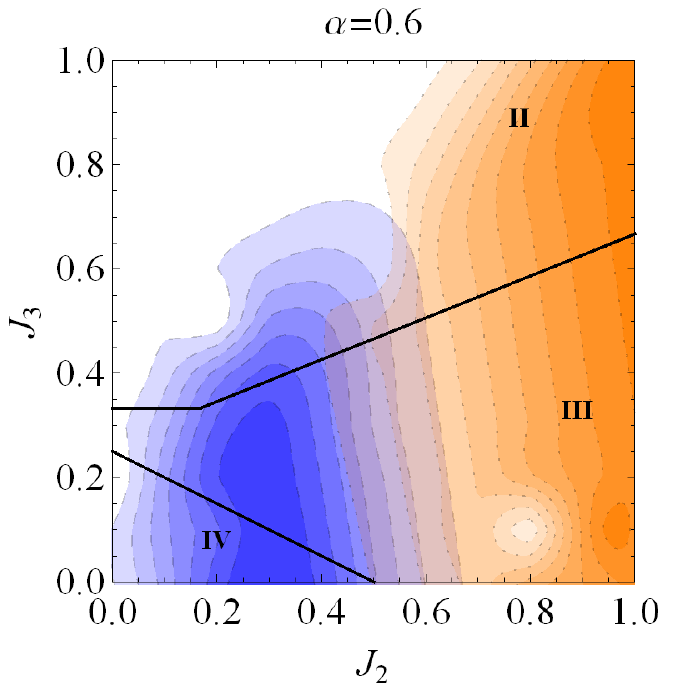}}
\subfigure[]{\includegraphics[height=125 pt]{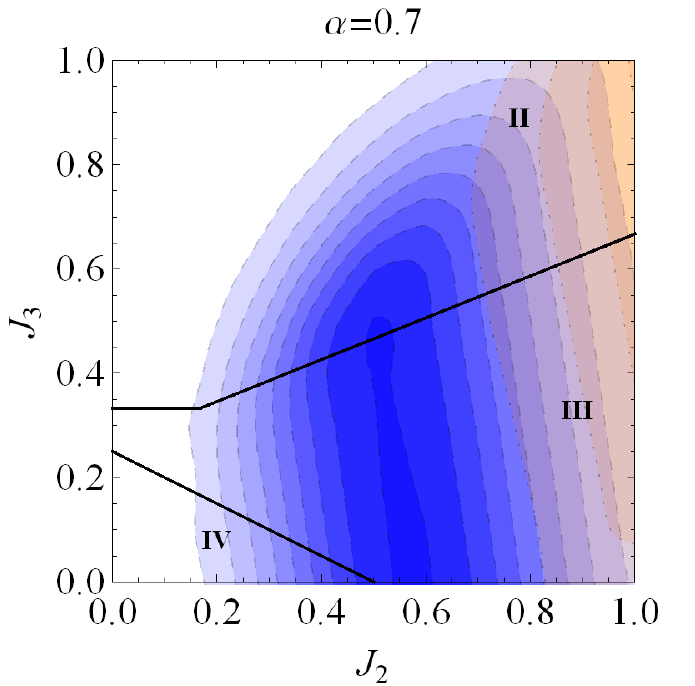}}
\caption{(Color online.) Fixed $\alpha$ slices in ($J_2,J_3$) showing the magnetically ordered phases (I, II, III, IV)=(\textit{Neel}, \textit{zigzag}, \textit{spiral}, \textit{stripy}) and shading corresponding to the ED $\chi(T)$ fit goodness. Increasing the Kitaev term (i.e.\ increasing $\alpha$) enlarges the extent of the \textit{zigzag} and \textit{stripy} phases, which occur at both small and large $\alpha$. Fits to $\text{Na}_2\text{Ir}\text{O}_3$ are shaded in orange (with dotted contour lines) and fits to $\text{Li}_2\text{Ir}\text{O}_3$ are in blue (with dashed contour lines); darker shading corresponds to good fitting with $\mu / \mu_B \approx 1$ and $J_2 \gtrsim J_3$, while lighter shading corresponds to poor agreement. Given a magnetically ordered ground state for each of the materials, the range of allowed parameters is found by intersecting the darker shaded region with the magnetically ordered phase. }
 \label{fig:slices}
\end{center}
\end{figure*}

In order to discuss results on the classical phase diagram we introduce standard nomenclature from the literature.  For each ordering wavevector the phases are labeled by a Roman numeral\cite{Albuqu,Fouet} as follows. 
 $\Gamma$: (I) \textit{Neel}. $M$: (IV) 	\textit{stripy}\cite{CJK}; and (II) \textit{zigzag}\cite{zigzag} (or \textit{columnar}\cite{OitmaaSingh}). All other wavevectors: (III) \textit{spiral}.
Figure \ref{fig:slices} displays six $(J_2,J_3)$ slices of the classical phase diagram at various fixed $\alpha$.

Quantum fluctuations modify the classical phase diagram in two ways. First, they create regions of quantum phases such as the plaquette valence bond solid or the Kitaev spin liquid; the former has been seen in the $J_1$-$J_2$-$J_3$ model\cite{Albuqu}, while the latter appears\cite{stationQ,CJK} at small $J_2, J_3$ starting at $\alpha \geq 0.8$. Second, they shift the boundaries between the magnetically ordered phases. Quantum fluctuations disfavor the spiral configurations\cite{Albuqu,Fouet} in favor of the collinear ordered phases, shrinking region (III); they also favor the \textit{Neel} state (I) over the other orders\cite{CJK,Albuqu}. 

The three dimensional phase diagram offers insights otherwise unavailable in its various limits. The \textit{stripy} (IV) region in the $J_1$-$J_2$-$J_3$ model at $\alpha=0$ is in the same phase as the fluctuation-free exactly solvable point $\alpha=0.5, J_2=J_3=0$ which may be understood only within the Kitaev-Heisenberg model\cite{CJK}. As $\alpha$ increases, both the \textit{stripy} and the \textit{zigzag} phases grow substantially larger. 
The dynamic frustration by the Kitaev term and the geometric frustration by the $J_2$ term have similar effects on the ordered phases, destabilizing \textit{Neel} in favor of \textit{stripy} and \textit{zigzag}.

It is worth reporting the direction of magnetic ordering in the various phases (excepting the special points $\alpha=0$ and $\alpha=1/2$). The direction of the collinear magnetic ordering in both \textit{stripy} and \textit{zigzag} phases is constrained already at the classical level. For $M_z$ \textit{stripy} order the spins lie along $S^z$, as was already determined by the spin rotation\cite{CJK} solution of the $J_2=J_3=0,\  \alpha=0.5$ Hamiltonian. For \textit{zigzag} order we found that the spins are constrained to the $S^x S^y$ plane (see Fig.\,\ref{fig:configs}).  Thermal and quantum fluctuations (``order from disorder'') force the spins to lie along a cubic axis within the classically allowed space, in this case the $S^x$ and $S^y$ axes. Trigonal distortion gives other perturbations: for example for $M_z$ \textit{stripy} order it cants the spin axis from $S^z$ toward the distortion axis, and for the \textit{Neel} phase the distortion axis may be an energy minimum or maximum within the Bloch sphere. A linear spin wave analysis found that directions closest to cubic axes are still preferred by quantum fluctuations. However, anisotropy in the real material likely overcomes all these effects to determine the ordering direction\cite{zigzag}.
	
Next we discuss the comparisons between experimentally measured susceptibility\cite{Na2IrO3,Li2IrO3} and exact diagonalization (ED), first describing each in turn. 
Uniform magnetic susceptibility data for the sodium and lithium materials at temperatures up to 300 K was taken from the most recent study\cite{Li2IrO3}, with the constant background removed\cite{Li2IrO3}. We used data from temperatures above 150K in order to avoid finite size effects when comparing to ED. ED using the ``fulldiag'' ALPS module\cite{ALPS} was performed keeping all eigenstates to enable comparison with high temperature data. The system diagonalized was an eight spin cluster, the unit cell of the $\alpha=1/2$ site dependent spin rotation\cite{CJK}, with periodic boundary conditions. As expected, the eight-spin ED, corresponding to a high temperature series expansion with eight-spin clusters, is reliable to far lower temperatures than the two-spin Curie-Weiss expression which only holds at $T \gg J$.	We found that ED finite size effects for eight-spin clusters were only visible in the susceptibility at low temperatures $T \lesssim J/2$, well below $J$.		The highest $J$ values needed for good fits were below the 150 K data cutoff, self consistently affirming the reliability of the ED fits.
	
	For each parameter set $(\alpha,J_2,J_3)$ we diagonalized the system to generate a curve $\chi(T)$. The Hamiltonian Eq.\,(\ref{eq:H}) with a magnetic field coupling term has two parameters in addition to $(\alpha,J_2,J_3)$, namely the overall scale $J$ and the magnetic field coupling $g \,\mu$. Since the effective spin-1/2 turns out to have the same $g$-factor as an electron spin, we fix $g=2$ and expect $\mu / \mu_B$ to remain close to $\mu / \mu_B =1$. For each $(\alpha,J_2,J_3)$ point the curve $\chi(T)$ was fit to the experimental data by the two parameters $J$ (corresponding to horizontal stretching) and $\mu / \mu_B$  (with $(\mu / \mu_B)^2$ corresponding to vertical stretching). The resulting fit was evaluated by a ``goodness function,'' the product of three Gaussian distributions, enforcing the following three conditions for a good fit. First, the magnetic moment $\mu / \mu_B$ found by the best fit must be close to 1, with a standard deviation of $0.15$. This constraint on $\mu$ effectively constrained $J$ as well. Second, the root-mean-square relative fit residual must be near zero with a standard deviation of $10^{-3}$. Third, the third neighbor coupling must be smaller or not much larger than the second neighbor coupling, $J_3 \lesssim J_2$, relaxed by a standard deviation of $0.2$. The absolute (unscaled) value of this goodness function was used to produce the shading in Fig.\,\ref{fig:slices}, with darker shading corresponding to better fits. 
	
	Given knowledge of the ground state magnetic order in $\text{Na}_2\text{Ir}\text{O}_3$ and $\text{Li}_2\text{Ir}\text{O}_3$, appropriate values for $\alpha$, $J_2$ and $J_3$ are found by intersecting the darker shaded regions in Fig.\,\ref{fig:slices} with the domain of the ordered phase. The estimated $\text{Na}_2\text{Ir}\text{O}_3$ and $\text{Li}_2\text{Ir}\text{O}_3$ parameters given either \textit{stripy} or \textit{zigzag} magnetic order are summarized in Table I. All material and order combinations yielded fitted values of $J$ in the range $J \approx $ 60--150 K, with the likeliest values $J \approx 100$ K. The lithium material has less structural distortion than the sodium material\cite{Li2IrO3}, suggesting a larger $\alpha$, in agreement with the fitting results if they have the same magnetic order. 
For zigzag ordered $\text{Li}_2\text{Ir}\text{O}_3$ we find $\alpha \approx $ 0.7, i.e.\ $J_K \sim $ 4--5 $J_1$ with a numerical value of $J_K \approx 130$ K. Such a large Kitaev term relative to the other couplings suggests that the Kitaev spin liquid phase may be within experimental reach\cite{Li2IrO3}. In particular, 
doping $\text{Li}_2\text{Ir}\text{O}_3$ may suppress its magnetic order to reveal characteristics of a doped Kitaev spin liquid\cite{topologicalSC}.

In conclusion, we propose the Kitaev-Heisenberg-$J_2$-$J_3$ model,
determining its ordered phases and further using ED fits of susceptibility measurements to demonstrate its applicability to Na$_2$IrO$_3$ and Li$_2$IrO$_3$. We find that the geometrical frustration due to $J_2,J_3$ and the dynamical frustration due to the Kitaev term both stabilize the same unconventional \textit{stripy} and \textit{zigzag} ordered ground states before the onset of the Kitaev spin liquid. 
We extract appropriate values for the spin couplings by first restricting to the experimentally observed magnetic order in the phase diagram, and then by requiring good fitting of the susceptibility $\chi(T)$ by ED data.
For \textit{zigzag} ordered Li$_2$IrO$_3$, a significant Kitaev term $J_K \approx 130$ K, five times larger than the nearest neighbor Heisenberg coupling, as well as substantial $J_2$ and $J_3$ couplings, are required for good agreement with experimental data.
\\

We are grateful to Ashvin Vishwanath, Tarun Grover and Hong Yao for valuable discussions and to Ashvin Vishwanath and Michael P. Zaletel for thoughtful comments on the manuscript.
 This work is supported in part by the National Science Foundation under Grant No. DGE 1106400 (I.K.), and by the China Scholarship Council and the Tsinghua Education Foundation in North America (Y.Z.Y.).
\begin{table}[htbp]
\begin{center}
\caption{\label{tab:params}Parameters for given $M$-wavevector order}
\begin{tabular}{c|c c c}
\hline\hline
 & \multicolumn{3}{c}{$\text{Na}_2\text{Ir}\text{O}_3$:} \\ \hline %\hline
 Stripy (IV) & $\alpha \approx $ 0.2--0.3,\ & $J_2 \lesssim 0.5$, $J_3\lesssim 0.2$ \ & $J\approx$ 110 K \\ \hline 
Zigzag (II) & $\alpha \approx $ 0.4--0.6, & $J_2,J_3 \gtrsim 0.4$ \ & $J\approx$ 100 K \\ \hline \hline
 & \multicolumn{3}{c}{$\text{Li}_2\text{Ir}\text{O}_3$:} \\ \hline %\hline
Stripy (IV) & $\alpha \approx 0.5$, & $J_2,J_3 \lesssim 0.3$  \ & $J\approx$ 100 K \\ \hline 
Zigzag (II) & $\alpha \approx $ 0.7, & $J_2,J_3 \gtrsim 0.4$  \ & $J\approx$ 90 K \\ \hline
%\hline 
\end{tabular}
\end{center}
\end{table}
%\vspace{-1mm} %for ease of printing

\end{document}